\documentclass[]{IEEEtran}
\usepackage{color}
\usepackage{graphicx}
\usepackage{mathptmx}
\onecolumn 
\begin{document}
\noindent	
{\bf \large Influence of autapses on synchronisation in
	neural networks with chemical synapses}\\
	
\noindent P. R. Protachevicz$^1$, K. C. Iarosz$^{2,3,*}$, I. L. Caldas$^1$,
C. G. Antonopoulos$^4$, A. M. Batista$^{1,5}$, J. Kurths$^{6,7,8}$\\
{\footnotesize \noindent 
$^1$Institute of Physics, University of S\~ao Paulo, S\~ao Paulo, SP, Brazil\\
$^2$Faculdade de Tel\^emaco Borba, FATEB, Tel\^emaco Borba, Paran\'a, Brazil\\
$^3$Graduate Program in Chemical Engineering, Federal University of Technology
Paran\'a, Ponta Grossa, Paran\'a, Brazil\\
$^4$Department of Mathematical Sciences, University of Essex, Wivenhoe Park, UK\\
$^5$Department of Mathematics and Statistics, State University of Ponta Grossa,
Ponta Grossa, PR, Brazil\\
$^6$Department Complexity Science, Potsdam Institute for Climate Impact
Research, Potsdam, Germany\\
$^7$Department of Physics, Humboldt University, Berlin, Germany\\
$^8$Centre for Analysis of Complex Systems, Sechenov First Moscow State Medical University, Moscow, Russia\\
}

\footnotesize Corresponding author: protachevicz@gmail.com.
		
\begin{abstract}  A great deal of research has been devoted on the investigation 
of neural dynamics in various network topologies. However, only a few studies 
have focused on the influence of autapses, synapses from a neuron onto itself via 
closed loops, on neural synchronisation. Here, we build a random network with 
adaptive exponential integrate-and-fire neurons coupled with chemical synapses, 
equipped with autapses, to study the effect of the latter on synchronous behaviour. 
We consider time delay in the conductance of the pre-synaptic neuron for excitatory 
and inhibitory connections. Interestingly, in neural networks consisting of both 
excitatory and inhibitory neurons, we uncover that synchronous behaviour depends 
on their synapse type. Our results provide evidence on the synchronous and 
desynchronous activities that emerge in random neural networks with chemical, 
inhibitory and excitatory synapses where neurons are equipped with autapses.
\end{abstract}
		
	\normalsize
	\section{Introduction}

An important research subject in neuroscience is to understand how cortical networks avoid or reach states of high synchronisation \cite{Kada2016}. In normal activity, excitatory and inhibitory currents are well balanced \cite{Tatti2018,Zhou2018}, while in epileptic seizures, high synchronous behaviour has been related to unbalanced current inputs \cite{Drongelen2005,Avoli2016}. Nazemi et al. \cite{Nazemi2018} showed that the structural coupling strength is important for the appearance of synchronised activities in excitatory and inhibitory neural populations. Various studies discuss the relation between structure and function in microscale and macroscale brain networks \cite{Sporns2013,DeBello2014,Sporns2016,Suarez2020}. In a microscale organisation, local excitatory and inhibitory connections are responsible for a wide range of neural interactions \cite{Sporns2012,Feng2018}. Bittner et al. \cite{Bittner2017} investigated population activity structure as a function of neuron types. They verified that the population activity structure depends on the ratio of excitatory to inhibitory neurons sampled. The pyramidal cell (excitatory neuron) exhibit spike adaptation, while the
fast spiking cell (inhibitory neuron) have a small or inexistent spike
adaptation \cite{Neske2015,Descalzo2005}.   

The excitatory to inhibitory and inhibitory to excitatory connections can change firing rates, persistent activities and synchronisation of the population of postsynaptic neurons \cite{Borgers2003,Han2018,Hayakawa2020,Kraynyukova2018,Mahmud2016}. Deco et al. \cite{Deco2014} analysed the effect of control in the inhibitory to excitatory coupling on the neural firing rate. Mejias et al. \cite{Mejias2016} proposed a computational model for the primary cortex in which different layers of excitatory and inhibitory connections were considered. 

A number of studies reported that excitatory synapses facilitate neural synchronisation \cite{Borges2017,Breakspear2003}, while inhibitory synapses have an opposite effect \cite{Kada2016,Ostojic2014,Protachevicz2019}. The time delay related to excitatory and inhibitory synapses influences the neural synchronisation \cite{Gu2015,Protachevicz2020}. Further on, there is a strong research interest in the investigation of how excitatory and inhibitory synapses influence synchronisation in neural networks \cite{Ge2019}. On the other hand, different types of networks have been used to analyse neural synchronisation, such as random \cite{Bondarenko1998,Gray2008}, small-world \cite{Antonopoulos2015,Antonopoulos2016,Hizanidis2016,Kim2013,Li2010,Qu2014}, regular \cite{Santos2019,Wang2007}, and scale-free \cite{Lombardi2017,Wang2011}.

Experiments showed that autapses are common in the brain and that they play an important role in neural activity \cite{Bekkers1998,Pouzat1998,Wang2015}. An autapse is a synaptic contact from a neuron to itself via a closed loop \cite{Bekkers2009,VanderLoos1972}, i.e. an auto-connection with a time delay on signal transmission \cite{Ergin2016}. Although, autaptic connections are anatomically present {\it in vivo} and in the neocortex, their functions are not completely understood \cite{Bacci2003}. Experimental and theoretical studies on excitatory and inhibitory autapses have been carried out \cite{Tamas1997,Saada-Madar2012,Suga2014,Szegedi2020} and the results have demonstrated that autaptic connections play a significant role in normal and abnormal brain dynamics \cite{Wyart2005,Valente2016,Wang2017,Yao2019}. The effects of autapses on neural dynamics were studied for single neurons \cite{Heng-Tong2015,Herrmann2004,Jia2018,Kim2019} and for neural networks \cite{HuiXin2014}. It has been shown that excitatory autapses contribute to a positive feedback \cite{Zhao2017} and can maintain persistent activities in neurons \cite{Bekkers2009}. It was also found that they promote burst firing patterns \cite{Wiles2017,Ke2019}. The inhibitory autapses contribute to a negative feedback \cite{Bacci2003,Zhao2017} and to the reduction of neural excitability \cite{Bekkers2003,Qin2014,Szegedi2020}. Guo et al. \cite{Guo2016} analysed chemical and electrical autapses in the regulation of irregular neural firing. In this way, autaptic currents can modulate neural firing rates \cite{Bacci2003}. Wang et al. \cite{Wang2014} demonstrated that chemical autapses can induce a filtering mechanism in random synaptic inputs. Interestingly, inhibitory autapses can favour synchronisation during cognitive activities \cite{Deleuze2019}. Short-term memory storage was observed by Seung et al. \cite{Seung2000} in a neuron with autapses submitted to excitatory and inhibitory currents. Finally, a study on epilepsy has exhibited that the number of autaptic connections can be different in her epileptic tissue \cite{Bacci2003}.

Here, we construct a random network with adaptive exponential integrate-and-fire (AEIF) neurons coupled with chemical synapses. The model of AEIF neurons was proposed by Brette and Gerstner \cite{Brette2005} and has been used to mimic neural spike and burst activities. Due to the fact that the chemical synapses can be excitatory and inhibitory, we build a network with excitatory synapses and autapses, a network with inhibitory synapses and autapses, and a network with both types of synapses and autapses. In the mixed network, we consider $80\%$ of excitatory and $20\%$ of inhibitory synapses and autapses. In this work, we focus on the investigation of the influence of autapses on neural synchronisation. Ladenbauer et al. \cite{Ladenbauer2013} studied the role of adaptation in excitatory and inhibitory populations of AEIF neurons upon synchronisation, depending on whether the recurrent synaptic excitatory or inhibitory couplings dominate. In our work, we show that not only the adaptation, but also the autapses can play an important role in the synchronous behaviour. To do so, we compute the order parameter to quantify synchronisation, the coefficient of variation in neural activity, firing rates and synaptic current inputs. In our simulations, we observe that autapses can increase or decrease synchronous behaviour in neural networks with excitatory synapses. However, when only inhibitory synapses are considered, synchronisation does not suffer significant alterations in the presence of autapses. Interestingly, in networks with excitatory and inhibitory synapses, we show that excitatory autapses can give rise to synchronous or desynchronous neural activity. Our results provide evidence how synchronous and desynchronous activities can emerge in neural networks due to autapses and contribute to understanding further the relation between autapses and neural synchronisation.

The paper is organised as follows: In Sec. \ref{sec_methods}, we introduce the neural network of AEIF neurons and the diagnostic tools that will be used, such as the order parameter for synchronisation, the coefficient of variation, the firing rates and synaptic current inputs. In Sec. \ref{sec_resutls_and_discussion}, we present the results of our study concerning the effects of autapses in neural synchronisation, and in Sec. \ref{sec_conclusions}, we draw our conclusions.

\section{Methods}\label{sec_methods}

\subsection{The AEIF model with neural autapses and network configurations}

The cortex comprises mainly excitatory pyramidal neurons and inhibitory interneurons \cite{Atencio2008}. Inhibitory neurons have a relatively higher firing rate than excitatory ones \cite{Wilson1994,Inawashiro1999,Baeg2001}. In the mammalian cortex, the firing pattern of excitatory neurons corresponds to regular spiking \cite{Neske2015}, while inhibitory neurons exhibit fast spiking activities \cite{Wang2016}. Furthermore, excitatory neurons show adaptation properties in response to depolarising inputs and the inhibitory adaptation current is negligible or nonexistent \cite{Foehring1991,Mancilla1998,Hensch2004,Destexhe2009,Masia2018,Borges2020}. The fast spiking interneurons are the most common inhibitory neurons in the cortex \cite{Puig2008}.

In the neural networks considered in this work, the dynamics of each neuron $j$, where $j=1,\ldots,N$, is given by the adaptive exponential integrate-and-fire model. In this framework, $N$ denotes the total number of neurons in the network. The AEIF model is able to reproduce different firing patterns, including regular and fast spiking \cite{Volo2019}. The network dynamics is given by the following set of coupled, nonlinear, ordinary differential equations
\begin{eqnarray}
	\label{eqIFrede}
	C_{\rm m}\frac{d V_j}{d t} & = & -g_{\rm L}(V_j-E_{\rm L})+g_{\rm L}{\Delta}_{\rm T} 
	\exp\left(\frac{V_j-V_{\rm T}}{{\Delta}_{\rm T}}\right)-w_j+I+I^{\rm chem}_j(t),
	\nonumber \\ 
	\tau_w\frac{dw_j}{dt} & = & a_j(V_j-E_{\rm L})-w_j,\\ 
	\tau_{\rm s} \frac{dg_{j}}{dt} & = &-g_{j}, \nonumber
\end{eqnarray}
where $V_j$ is the membrane potential, $w_j$ the adaptation current and $g_j$ the synaptic conductance of neuron $j$. $k$ and $j$ identify the pre and postsynaptic neurons. When the membrane potential of neuron $j$ is above the threshold $V_{\rm thres}$, i.e. when $V_j>V_{\rm thres}$ \cite{Naud2008}, the state variables are updated according to the rules
\begin{eqnarray}\label{eq_variables_update}
	V_j & \to &V_{\rm r}, \nonumber\\
	w_j & \to& w_j + b_j,\\
	g_{j} & \to &g_{j}+g_{\rm s}, \nonumber
\end{eqnarray}
where $g_{\rm s}$ assumes the value $g_{\rm e}^{\rm aut}$ for excitatory autapses, $g_{\rm e}$ for synapses among excitatory neurons, $g_{\rm ei}$ for synapses from excitatory to inhibitory neurons, $g_{\rm i}^{\rm aut}$ for inhibitory autapses, $g_{\rm i}$ for synapses among inhibitory neurons and $g_{\rm ie}$ for synapses from inhibitory to excitatory neurons. We consider a neuron is excitatory (inhibitory) when it is connected to another neuron with an excitatory (inhibitory) synapse. The initial conditions of $V_j$ are randomly distributed in the interval $V_j=[-70,-50]$ mV. The initial values of $w_j$ are randomly distributed in the interval $w_j=[0,300]$ pA for excitatory and $w_j=[0,80]$ pA for inhibitory neurons. We consider the initial value of $g_j$ equal to zero for all neurons. Table \ref{table1} summarises the description and values of the parameters used in the simulations.

\begin{table}[htb]
	\caption{Description and values of the parameters in the AEIF system \ref{eqIFrede} and \ref{eq_variables_update} used in the simulations. Values for parameters for excitatory and inhibitory connections are denoted by $\bullet$ and $\star$, respectively.}\label{table1}
	\begin{center}
		\begin{tabular}{c l l}
			\hline
			\small Parameter & \small Description &\small Value \\ 	
			\hline
			\small $N$ &\footnotesize Number of AEIF neurons & \small 1000 neurons \\ 
			\small $C_{\rm m}$ &\footnotesize Membrane capacitance &\small 200 pF \\ 
			\small $g_{\rm L}$ &\footnotesize Leak conductance & \small 12 nS \\ 
			\small $E_{\rm L}$ &\footnotesize Leak reversal potential &\small -70 mV \\ 
			\small $I$ &\footnotesize Constant input current & \small 270 pA \\ 
			\small $\Delta_{\rm T}$	&\footnotesize Slope factor &\small 2 mV \\
			\small $V_{\rm T}$	&\footnotesize Potential threshold & \small-50 mV \\ 
			\small $\tau_w$ & \footnotesize Adaptation time constant & \small 300 ms\\
			\small $\tau_{\rm s}$ & \footnotesize Synaptic time constant &\small 2.728 ms \\	
			\small $V_{\rm r}$ &\footnotesize Reset potential & \small -58 mV \\ 
			\small $M_{jk}^{\rm exc}$ &\footnotesize Adjacency matrix elements &\small 0 or 1 \\
			\small $M_{jk}^{\rm inh}$ &\footnotesize Adjacency matrix elements &\small 0 or 1 \\
			\small $t_{\rm ini}$ &\footnotesize Initial time in the analyses &\small 10 s \\ 
			\small $t_{\rm fin}$ &\footnotesize Final time in the analyses &\small 20 s \\ 
			\small $a_j$ &\footnotesize Subthreshold adaptation &\small $[1.9,2.1]$ nS $\bullet$ \\ 
			\small & &\small 0 nS $\star$ \\ 
			\small $b_j$ &\footnotesize Triggered adaptation &\small 70 pA $\bullet$ \\ 
			\small & & \small 0 pA $\star$\\ 
			\small $V_{\rm REV}$&\footnotesize Synaptic reversal potential &\small 
			\small $V_{\rm REV}^{\rm exc}=0$ mV $\bullet$ \\ 
			\small & &\small $V_{\rm REV}^{\rm inh}=-80$ mV $\star$ \\ 
			\small $g_{\rm s}$ &\footnotesize Chemical conductances &\small $g_{\rm e}$,
			$g_{\rm e}^{\rm aut}$, $g_{\rm ei}$ $\bullet$ \\	
			\small & &\small $g_{\rm i}$, $g_{\rm i}^{\rm aut}$, $g_{\rm ie}$ $\star$ \\	
			\small $g_{\rm e}$ &\footnotesize Excitatory to excitatory &\small  [0,0.5] nS $\bullet$ \\
			\small $g_{\rm e}^{\rm aut}$ &\footnotesize Excitatory autaptic &\small  [0,35] nS $\bullet$ \\
			\small $g_{\rm ei}$ &\footnotesize Excitatory to inhibitory &\small  [0,5] nS $\bullet$ \\
			\small $g_{\rm i}$ &\footnotesize Inhibitory to inhibitory &\small  [0,2] nS $\star$ \\ 
			\small $g_{\rm i}^{\rm aut}$ &\footnotesize Inhibitory autaptic &\small  [0,100] nS $\star$ \\
			\small $g_{\rm ie}$ &\footnotesize Inhibitory to excitatory &\small  [0,3] nS $\star$ \\
			\small $d_j$ &\footnotesize Time delay &\small $d_{\rm exc}=1.5$ ms $\bullet$ \\
			\small & &\small $d_{\rm inh}=0.8$ ms $\star$ \\
			\hline 
		\end{tabular}
	\end{center}
\end{table}

The synaptic current arriving at each neuron depends on specific parameters, including the connectivity encoded in the adjacency matrices $M^{\rm exc}$ and $M^{\rm inh}$, i.e. in the excitatory and inhibitory connectivity matrices. In particular, the input current $I^{\rm chem}_j$ arriving at each neuron $j$, is calculated by
\begin{equation}\label{eq:I_j_syn} 
I^{\rm chem}_j(t)=I^{\rm exc}_j(t)+I^{\rm inh}_j(t),
\end{equation}
where
\begin{eqnarray}
	I^{\rm exc}_j(t) & =& I^{\rm ee}_j(t)+I^{\rm ei}_j(t)+I^{\rm e,aut}_j(t)\nonumber \\ 
	& =& (V_{\rm{REV}}^{\rm exc}-V_j(t)) \sum_{k=1}^{N}M_{jk}^{\rm exc} g_{k}(t-d_{\rm exc})
\end{eqnarray}
and 
\begin{eqnarray}
	I^{\rm inh}_j(t) & =& I^{\rm ii}_j(t)+I^{\rm ie}_j(t)+I^{\rm i,aut}_j(t)\nonumber\\ 
	& = &(V_{\rm{REV}}^{\rm inh}-V_j(t)) \sum_{k=1}^{N}M_{jk}^{\rm inh}g_{k}(t-d_{\rm inh}).
\end{eqnarray}
In this framework, the type of synapse (excitatory or inhibitory) depends on the synaptic reversal potential $V_{\rm{REV}}$. We consider $V_{\rm REV}^{\rm exc}=0$ mV for excitatory and $V_{\rm REV}^{\rm inh}=-80$ mV for inhibitory synapses. The time delay in the conductance of the pre-synaptic neuron $k$ ($g_k$) assumes $d_{\rm exc}=1.5$ ms for excitatory and $d_{\rm inh}=0.8$ ms for inhibitory connections \cite{Borges2020}. The influence of delayed conductance on neural synchronisation was studied in \cite{Protachevicz2020}. There are no spike activities in the time interval $t=[-d_j,0]$.

\begin{figure}[hbt]
	\centering
	\includegraphics[scale=0.15]{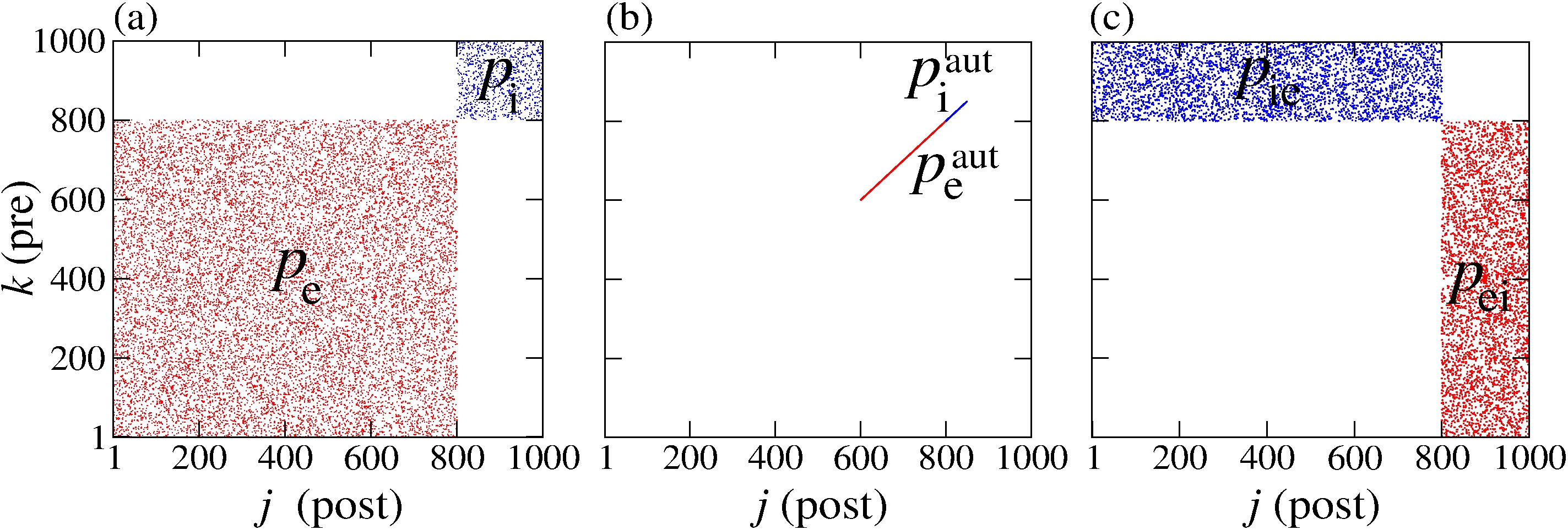}
	\caption{(Colour online) Representation of the connections: (a) in the same population, (b) for autapses and (c) among different neural populations. Here, ``pre'' stands for ``pre-synaptic'' and ``post'' for ``post-synaptic''. We note that we have used $P_{\rm exc}=80\%$ excitatory (denoted red) and $P_{\rm inh}=20\%$ inhibitory (denoted blue) neural populations which amounts to a total of $N=1000$ neurons.}\label{fig1}
\end{figure}

The first $N_{\rm exc}$ neurons are excitatory and the last $N_{\rm inh}$ inhibitory. The connections that depart from excitatory and inhibitory neurons are associated with the excitatory and inhibitory matrices, $M^{\rm exc}$ and $M^{\rm inh}$, where each entry is denoted $M_{jk}^{\rm exc}$ and $M_{jk}^{\rm inh}$, respectively. These adjacency matrices are binary and have entries equal to 1 when there is a connection from neuron $k$ to neuron $j$, or 0 otherwise, as shown in Fig. \ref{fig1}.

We consider $P_{\rm exc}=80\%$ excitatory and $P_{\rm inh}=20\%$ inhibitory neural populations following \cite{Volo2019,Noback2005}, where the numbers of excitatory and inhibitory neurons are given by $N_{\rm exc}=P_{\rm exc}N$ and $N_{\rm inh}=P_{\rm inh}N$, respectively. The connectivity probabilities are set to $p_{\rm e}^{\rm aut}=p_{\rm i}^{\rm aut}=0.25$ for excitatory and inhibitory autapses, to $p_{\rm e}=0.05$ and $p_{\rm i}=0.2$ for connectivity within the same neural population and to $p_{\rm ei}=p_{\rm ie}=0.05$ for connectivity among different neural populations \cite{Volo2019}. The subscripts ``e'' and ``i'' stand for ``excitatory'' and ``inhibitory'', respectively and the superscript ``aut'' stands for ``autapses''. The terms $p_{\rm ei}$ and $p_{\rm ie}$ represent the probabilities of connections from excitatory to inhibitory and from inhibitory to excitatory neurons, respectively.

The probabilities of excitatory and inhibitory autapses are defined by
\begin{eqnarray}
	p_{\rm e}^{\rm aut}=\frac{N_{\rm e}^{\rm aut}}{N_{\rm exc}}\quad{\rm and}\quad p_{\rm i}^{\rm aut}= \frac{N_{\rm i}^{\rm aut}}{N_{\rm inh}}, 
\end{eqnarray}
where $N_{\rm e}^{\rm aut}$ and $N_{\rm i}^{\rm aut}$ are the number of autapses in the excitatory and inhibitory populations, respectively. For a network with only excitatory (inhibitory) neurons, the number of excitatory (inhibitory) neurons is $N_{\rm exc}=N$ ($N_{\rm inh}=N$). For connections within the excitatory and inhibitory populations, the corresponding probabilities $p_{\rm e}$ and $p_{\rm i}$ are given by
\begin{eqnarray}
	p_{\rm e}=\frac{N_{\rm e}}{N_{\rm exc}(N_{\rm exc}-1)} \quad{\rm and}\quad p_{\rm i}= \frac{N_{\rm i}}{N_{\rm inh}(N_{\rm inh}-1)}, 
\end{eqnarray}
where $N_{\rm e}$ and $N_{\rm i}$ are the number of synaptic connections in the excitatory and inhibitory populations, respectively. For connections among different populations, the corresponding probabilities are given by
\begin{eqnarray}
	p_{\rm ei}=\frac{N_{\rm ei}}{N_{\rm exc}N_{\rm inh}} \quad{\rm and}\quad p_{\rm ie}=\frac{N_{\rm ie}}{N_{\rm exc}N_{\rm inh}},
\end{eqnarray}
where $N_{\rm ei}$ and $N_{\rm ie}$ are the number of synaptic connections from the excitatory to the inhibitory and from the inhibitory to the excitatory populations, respectively. Therefore, when only one neural population is considered, $p_{\rm ei}$ and $p_{\rm ie}$ cannot be defined. The resulting 6 connectivity probabilities are represented in the connectivity matrix in Fig. \ref{fig1}, where $k$ and $j$ denote the pre- and post-synaptic neurons, respectively. \mbox{Figure \ref{fig1}} shows the connections associated to probabilities: (a) in the same population ($p_{\rm e}$ and $p_{\rm i}$), (b) for autapses ($p_{\rm e}^{\rm aut}$ and $p_{\rm i}^{\rm aut}$) and (c) among different populations ($p_{\rm ei}$ and $p_{\rm ie}$).

Finally, we associate the conductances $g_{\rm e}$, $g_{\rm i}$, $g_{\rm e}^{\rm aut}$, $g_{\rm i}^{\rm aut}$, $g_{\rm ei}$ and $g_{\rm ie}$ to the corresponding connectivity probabilities discussed before. To solve the set of ordinary differential equations in system \ref{eqIFrede}, we used the 4th order Runge-Kutta method with the integration time-step equal to $10^{-2}$ ms.

\vspace{1cm}
\subsection{Computation of neural synchronisation}

Synchronous behaviour in neural networks can be quantified by means of the order parameter $R$ \cite{Kuramoto1984}
\begin{eqnarray}
	R(t)=\Bigg|\frac{1}{N}\sum_{j=1}^{N}\exp\big({\rm i}\psi_{j}(t)\big)\Bigg|,
\end{eqnarray}
where $R(t)$ is the amplitude of a centroid phase vector over time, ${\rm i}$ the imaginary unit, satisfying ${\rm i}^2=-1$ and $|\cdot|$, the vector-norm of the argument. The phase of each neuron $j$ in time is obtained by means of
\begin{eqnarray}\label{eq_neuron_phase}
\psi_{j}(t)=2\pi m+2\pi\frac{t-t_{j,m}}{t_{j,m+1}-t_{j,m}},
\end{eqnarray}
where $t_{j,m}$ is the time of the $m$-th spike of neuron $j$, where $t_{j,m}<t<t_{j,m+1}$ \cite{Rosenblum97}. We consider that spikes occur whenever $V_j>V_{\rm thres}$ \cite{Naud2008}. $R(t)$ takes values in $[0,1]$ and, is equal to $0$ for completely desynchronised neural activity and $1$ for fully synchronised neural behaviour. We compute the time-average order parameter $\overline{R}$ \cite{Batista2017}, given by 
\begin{eqnarray}\label{ref:Rmedio}
\overline{R}=\frac{1}{t_{\rm fin}- t_{\rm ini}} \int_{t_{\rm ini}}^{t_{\rm fin}} R(t)dt,
\end{eqnarray} 
where $(t_{\rm fin}-t_{\rm ini})$ is the length of the time window $[t_{\rm ini},t_{\rm fin}]$. Here, we have used $t_{\rm ini}=10$ s and $t_{\rm fin}=20$ s. Similarly, we calculate the synchronisation of the non-autaptic neurons
\begin{eqnarray}
	R_{\rm non}(t)=\Bigg|\frac{1}{N^{\rm non}}\sum_{j=1}^{N^{\rm non}}\exp\Big({\rm i}\psi_{j}^{\rm non}(t)\Big)\Bigg|
\end{eqnarray}
and autaptic neurons
\begin{eqnarray}
	R_{\rm aut}(t)=\Bigg|\frac{1}{N^{\rm aut}}\sum_{j=1}^{N^{\rm aut}}\exp\Big({\rm i}\psi_{j}^{\rm aut}(t)\Big)\Bigg|,
\end{eqnarray}
where $N^{\rm non}$ and $N^{\rm aut}$ are the number of non-autaptic and autaptic neurons, respectively. In this context, $\psi_{j}^{\rm non}$ and $\psi_{j}^{\rm aut}$ are the phases of the non-autaptic and autaptic neuron $j$ and both terms are computed using Eq. \ref{eq_neuron_phase} for the times of spiking of the non-autaptic and autaptic neurons, respectively. $\overline{R}_{\rm non}$ and $\overline{R}_{\rm aut}$ are then obtained according to Eq. \ref{ref:Rmedio}.

\subsection{Mean coefficient of variation of interspike intervals}

We calculate the interspike intervals of each neuron to obtain the mean coefficient of variation. In particular, the $m$-th interspike interval of neuron $j$, ${\rm ISI}_j^m$, is defined as the difference between two consecutive spikes,
\begin{eqnarray}
	{\rm ISI}_j^m=t_{j,m+1}-t_{j,m}>0,
\end{eqnarray}
where $t_{j,m}$ is the time of the $m$-th spike of neuron $j$. Using the mean value of ${\rm ISI}_j$ over all $m$, $\overline{\rm ISI}_j$ and its standard deviation $\sigma_{{\rm ISI}_j}$, we can compute the coefficient of variation (CV) of neuron $j$,
\begin{eqnarray}
	{\rm CV}_j = \frac{\sigma_{{\rm ISI}_j}}{\overline{\rm ISI}_j}.
\end{eqnarray}
The average ${\rm CV}$ over all neurons in the network, $\overline{\rm CV}$, can then be computed by
\begin{eqnarray}
	\overline{\rm CV}=\frac{1}{N}\sum_{j=1}^{N} {\rm CV}_j.
\end{eqnarray}
We use the value of $\overline{\rm CV}$ to identify spikes whenever $\overline{\rm CV}<0.5$ and burst firing patterns whenever $\overline{\rm CV}\ge0.5$ \cite{Borges2017,Protachevicz2018} in neural activity.

\subsection{Firing rates in neural populations}

The mean firing-rate of all neurons in a network is computed by means of
\begin{eqnarray}
	\overline{F}=\frac{1}{N(t_{\rm fin}-t_{\rm ini})}\sum_{j=1}^{N} 
	\Bigg(\int_{t_{\rm ini}}^{t_{\rm fin}}\delta(t'-t_j)dt'\Bigg),
\end{eqnarray}
where $t_j$ is the firing time of neuron $j$. In some occasions, we calculate the mean firing frequency of neurons with and without autapses,
\begin{eqnarray}
	\overline{F}_{\rm aut}=\frac{1}{N_{\rm x}^{\rm aut}(t_{\rm fin}-t_{\rm ini})}
	\sum_{j=1}^{N_{\rm x}^{\rm aut}}\Bigg(\int_{t_{\rm ini}}^{t_{\rm fin}}\delta(t'-t_j^{\rm aut})dt'
	\Bigg)
\end{eqnarray}
and 
\begin{eqnarray}
	\overline{F}_{\rm non}=\frac{1}{N_{\rm x}^{\rm non}(t_{\rm fin}-t_{\rm ini})}
	\sum_{j=1}^{N_{\rm x}^{\rm non}}\Bigg(\int_{t_{\rm ini}}^{t_{\rm fin}}\delta(t'
	-t_j^{\rm non})dt'\Bigg),
\end{eqnarray}
where $N_{\rm x}^{\rm aut}$ and $ N_{\rm x}^{\rm non}$ are the number of neurons with and without autapses, and $t_j^{\rm aut}$ and $t_j^{\rm non}$ the firing times of neurons with and without autapses. The subscript ``x'' denotes the population of excitatory (``e'') or inhibitory (``i'') neurons.

Similarly, we calculate the firing rate of excitatory and inhibitory neurons by means of
\begin{eqnarray}
	\overline{F}_{\rm exc}=\frac{1}{N_{\rm exc}(t_{\rm fin}-t_{\rm ini})}
	\sum_{j=1}^{N_{\rm exc}}\left(\int_{t_{\rm ini}}^{t_{\rm fin}}\delta(t'-t_j^{\rm exc})dt'
	\right)
\end{eqnarray}
and
\begin{eqnarray}
	\overline{F}_{\rm inh}=\frac{1}{N_{\rm inh}(t_{\rm fin}-t_{\rm ini})}
	\sum_{j=1}^{N_{\rm inh}}\left(\int_{t_{\rm ini}}^{t_{\rm fin}}\delta(t'-t_j^{\rm inh})dt'
	\right),
\end{eqnarray}
where $t_j^{\rm exc}$ and $t_j^{\rm inh}$ are the firing times of the excitatory and inhibitory neurons, respectively.

\vspace{1cm}
\subsection{Synaptic current inputs}

In our work, we calculate the mean instantaneous input $I^{\rm chem}(t)$ and the time average of the synaptic input $\overline{I}_{\rm s}$ (pA) in the network by
\begin{eqnarray}
	I^{\rm chem}(t)=\frac{1}{N}\sum_{j=1}^{N}I_j^{\rm chem}(t)
\end{eqnarray}
and
\begin{eqnarray}
	\overline{I}_{\rm s}=\frac{1}{t_{\rm fin}-t_{\rm ini}}\int_{t_{\rm ini}}^{t_{\rm fin}}
	I^{\rm chem}(t) dt,
\end{eqnarray} 	
respectively, where $I_j^{\rm chem}(t)$ is given by Eq. \ref{eq:I_j_syn}. In this respect, the values of $I^{\rm chem}$ change over time due to excitatory and inhibitory inputs received by neuron $j$, where $j=1,\ldots,N$.


\section{Results and Discussion}\label{sec_resutls_and_discussion}

\subsection{Network with excitatory neurons only}

Networks with excitatory neurons were studied previously by Borges et al. \cite{Borges2017} and Protachevicz et al. \cite{Protachevicz2019}. These studies showed that excitatory neurons can change firing patterns and improve neural synchronisation. Fardet et al. \cite{Fardet2018} and Yin et al. \cite{Yin2018} reported that excitatory autapses with few milliseconds time delay can change neural activities from spikes to bursts. Wiles et. al \cite{Wiles2017} demonstrated that excitatory autaptic connections contribute more to bursting firing patterns than inhibitory ones.

\begin{figure}[hbt]
	\centering
	\includegraphics[scale=0.18]{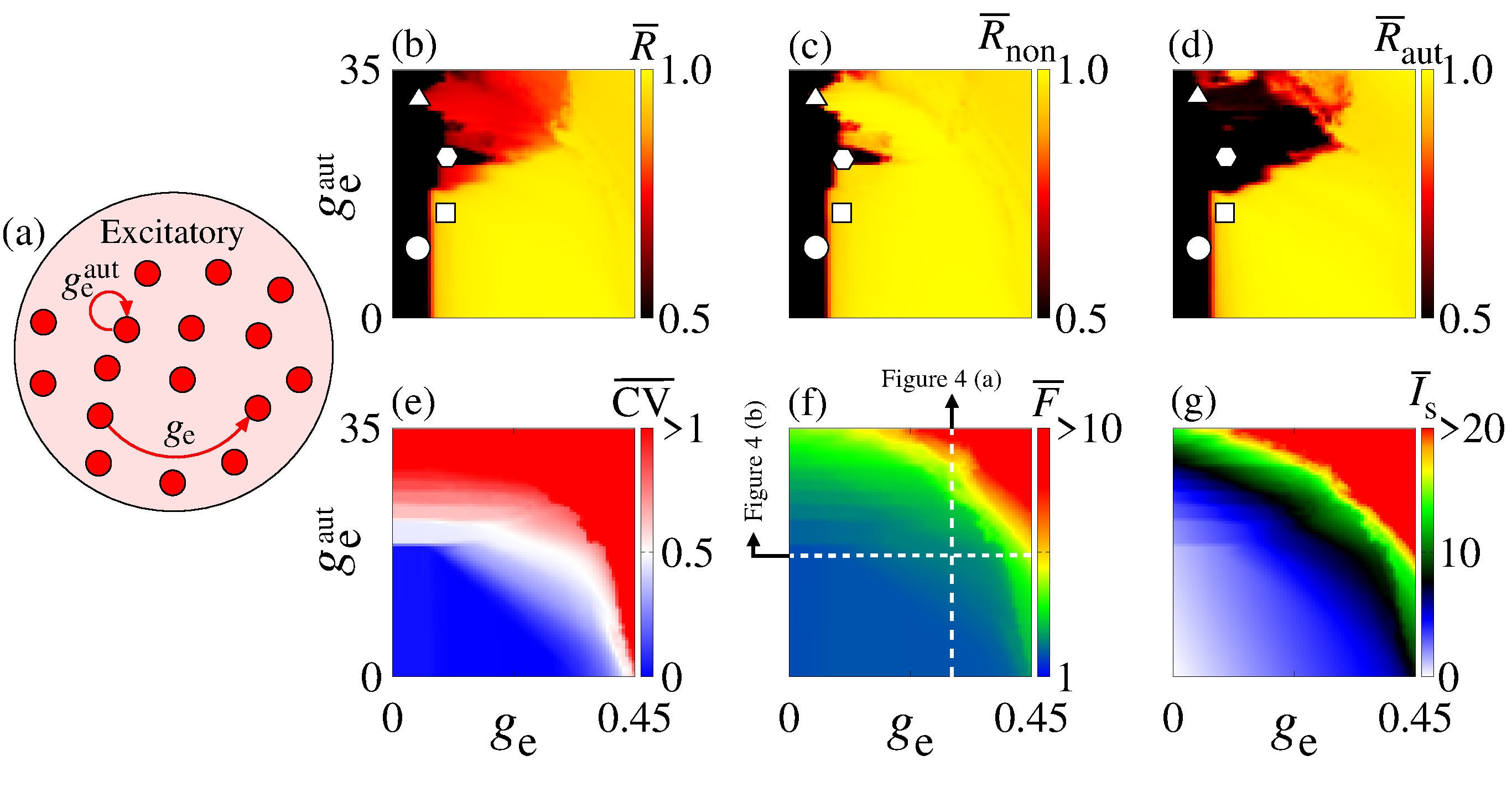}
	\caption{(Colour online) (a) Schematic representation of the neural network where $g_{\rm e}$ is the intensity of excitatory synaptic conductance and $g_{\rm e}^{\rm aut}$ of the excitatory autaptic conductance. Parameter space $g_{\rm e}\times g_{\rm e}^{\rm aut}$, where the colour bars correspond to (b) $\overline{R}$, (c) $\overline{R}_{\rm non}$, (d) $\overline{R}_{\rm aut}$, (e) $\overline{\rm CV}$, (f) $\overline{F}$ and (g) $\overline{I}_{\rm s}$. The raster plots of the parameters indicated in (b), (c) and (d) (circle, square, triangle and hexagon) are shown in Fig. \ref{fig3}. The vertical and horizontal white, dash, lines in panel (f) are used to vary $g_{\rm e}^{\rm aut}$ and $g_{\rm e}$ in the computations in panels (a) and (b) in Fig. \ref{fig4}, respectively. The closed loop in panel (a) corresponds to an autapse of excitatory autaptic conductance $g_{\rm e}^{\rm aut}$.}\label{fig2}
\end{figure}

\begin{figure}[hbt]
	\centering
	\includegraphics[scale=0.2]{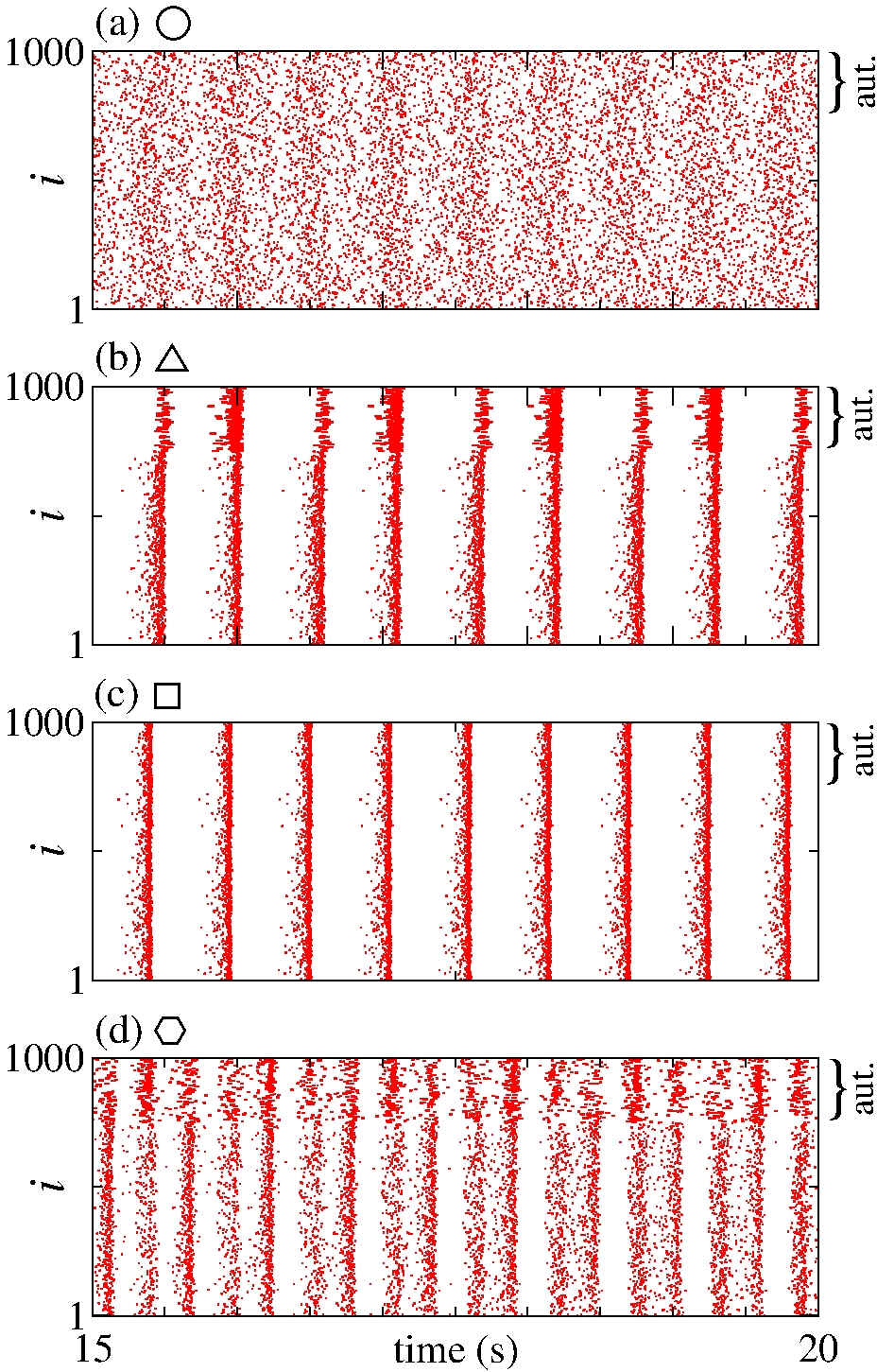}
	\caption{(Colour online) Raster plots for the neural network with excitatory neurons only. The values of the parameters $g_{\rm e}$ and $g_{\rm e}^{\rm aut}$ are indicated in panels (b) to (d) in Fig. \ref{fig2} by circle, triangle, square and hexagon symbols, respectively. The curly brackets in the upper right corner of the plots denote the autaptic neurons considered.}\label{fig3}
\end{figure}

In Fig. \ref{fig2}, we consider a neural network with excitatory neurons only, where $g_{\rm e}$ corresponds to the intensity of excitatory synaptic conductance and $g_{\rm e}^{\rm aut}$ to the intensity of excitatory autaptic conductance. In our neural network, a neuron receives many connections from other neurons with small intensity of synaptic conductances. For the autaptic neurons, only one synaptic contact from a neuron to itself via a closed loop is considered. Due to this fact, to study the autaptic influence on the high synchronous activities, we consider values of $g_{\rm e}^{\rm aut}$ greater than $g_{\rm e}$. Panel (a) shows a schematic representation of a neural network of excitatory neurons only with a single autapse represented by the closed loop with excitatory autaptic conductance $g_{\rm e}^{\rm aut}$. Panels (b) to (d) give the mean order parameter in the parameter space $g_{\rm e}\times g_{\rm e}^{\rm aut}$. We see that excitatory autapses can increase or reduce the synchronisation in a population of excitatory neurons when the intensity of the excitatory synaptic conductance is small. In these panels, the circle ($g_{\rm e}=0.05$ nS and $g_{\rm e}^{\rm aut}=10$ nS), triangle \mbox{($g_{\rm e}=0.05$ nS} and $g_{\rm e}^{\rm aut}=31$ nS), square ($g_{\rm e}=0.1$ nS and $g_{\rm e}^{\rm aut}=15$ nS) and hexagon ($g_{\rm e}=0.1$ nS and \mbox{$g_{\rm e}^{\rm aut}=22$ nS)} symbols indicate the values of the parameters shown in Fig. \ref{fig3}. We observe that desynchronous firing patterns as seen in Fig. \ref{fig3}(a) can become more synchronous, as it can be seen in Fig. \ref{fig3}(b), due the increase of the excitatory autaptic conductance. On the other hand, the increase of the autaptic conductance can decrease the level of synchronisation in the network, i.e. from high in Fig. \ref{fig3}(c) to low synchronous activities in Fig. \ref{fig3}(d). However, as shown in Fig. \ref{fig2}(d), the autaptic connections affect mainly the synchronisation of autaptic neurons. 

For a strong excitatory synaptic coupling ($g_{\rm e}\ge 0.3$), autapses do not reduce neural synchronisation significantly. Panels (e) to (g) in Fig. \ref{fig2} show the mean coefficient of variation ($\overline{\rm CV}$), firing frequency ($\overline{F}$) and synaptic current ($\overline{I}_{\rm s}$), respectively. We verify that the excitatory autaptic neurons promote the increase of $\overline{\rm CV}$, $\overline{F}$ and $\overline{I}_{\rm s}$ in the network. In Fig. \ref{fig2}(e), we find that both synaptic and autaptic couplings can lead to burst activities, as reported by Borges et al. in \cite{Borges2017} and Fardet et al. in \cite{Fardet2018}. The burst and spike activities are characterised by $\overline{\rm CV}\ge 0.5$ (red region) and $\overline{\rm CV}<0.5$ (blue region), respectively. In addition, excitatory autaptic neurons can change the firing patterns of all neurons in the network from spike to burst activities. In panels (f) and (g) in Fig. \ref{fig2}, we observe that excitatory autapses contribute to the increase of the mean firing frequency and synaptic current.

\begin{figure}[hbt]
	\centering
	\includegraphics[scale=0.2]{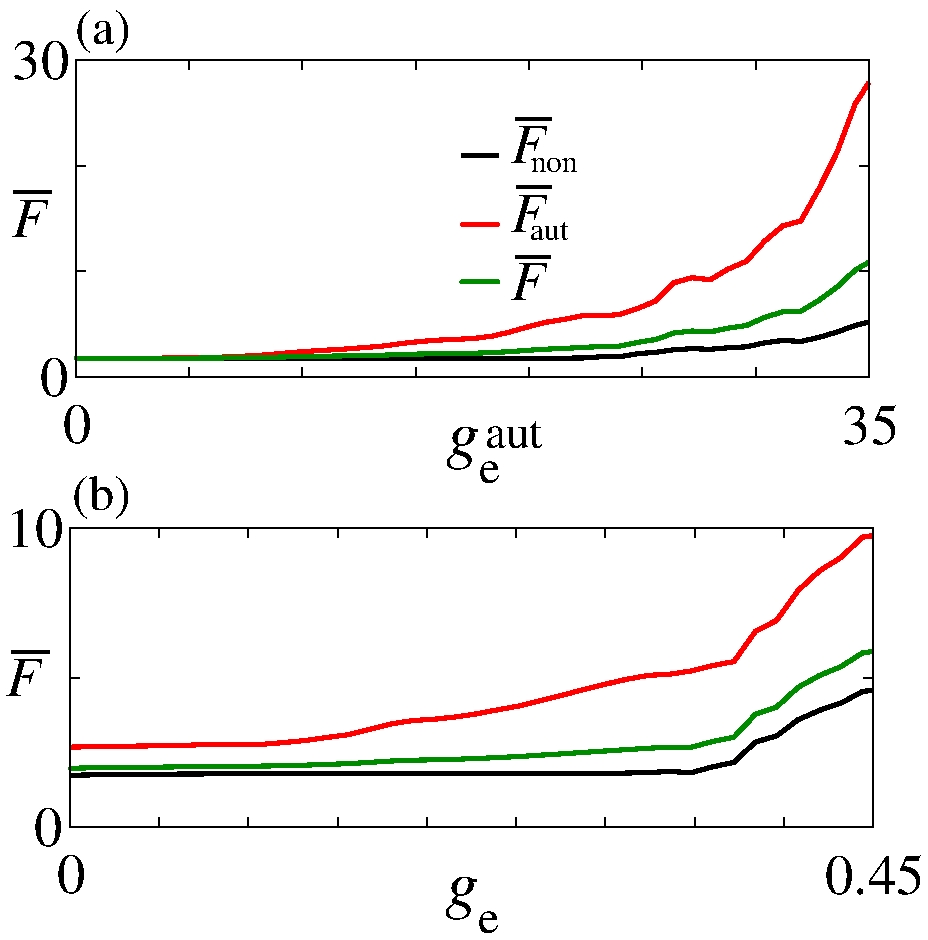}
	\caption{(Colour online) Plot of $\overline{F}_{\rm non}$ (black curve), $\overline{F}_{\rm aut}$ (red curve) and $\overline{F}$ (green curve) for \mbox{(a) $g_{\rm e}=0.3$ nS} varying $g_{\rm e}^{\rm aut}$ and (b) $g_{\rm e}^{\rm aut}=20$ nS varying $g_{\rm e}$. Here, $g_{\rm e}^{\rm aut}$ and $g_{\rm e}$ vary along the white, dash, lines in \mbox{Fig. \ref{fig2}(f).}} 
	\label{fig4}
\end{figure}

Next, we analyse the influence of autaptic connections on neural firing frequency. Figure \ref{fig4} shows the mean firing frequency of neurons without ($\overline{F}_{\rm non}$) and with autapses ($\overline{F}_{\rm aut}$), as well as of all neurons in the excitatory network ($\overline{F}$). In Fig. \ref{fig4}(a), we consider $g_{\rm e}=0.3$ nS varying $g_{\rm e}^{\rm aut}$, while in Fig. \ref{fig4}(b), we use $g_{\rm e}^{\rm aut}=20$ nS varying $g_{\rm e}$, as shown in Fig. \ref{fig2}(f) with white, dash, lines. We find that the autaptic connections increase the firing frequency of all neurons in the network and mainly those with autaptic connections. In our simulations, neurons with excitatory autapses exhibit the highest firing rate.

\subsection{Network with inhibitory neurons only}

\begin{figure}[hbt]
	\centering
	\includegraphics[scale=0.18]{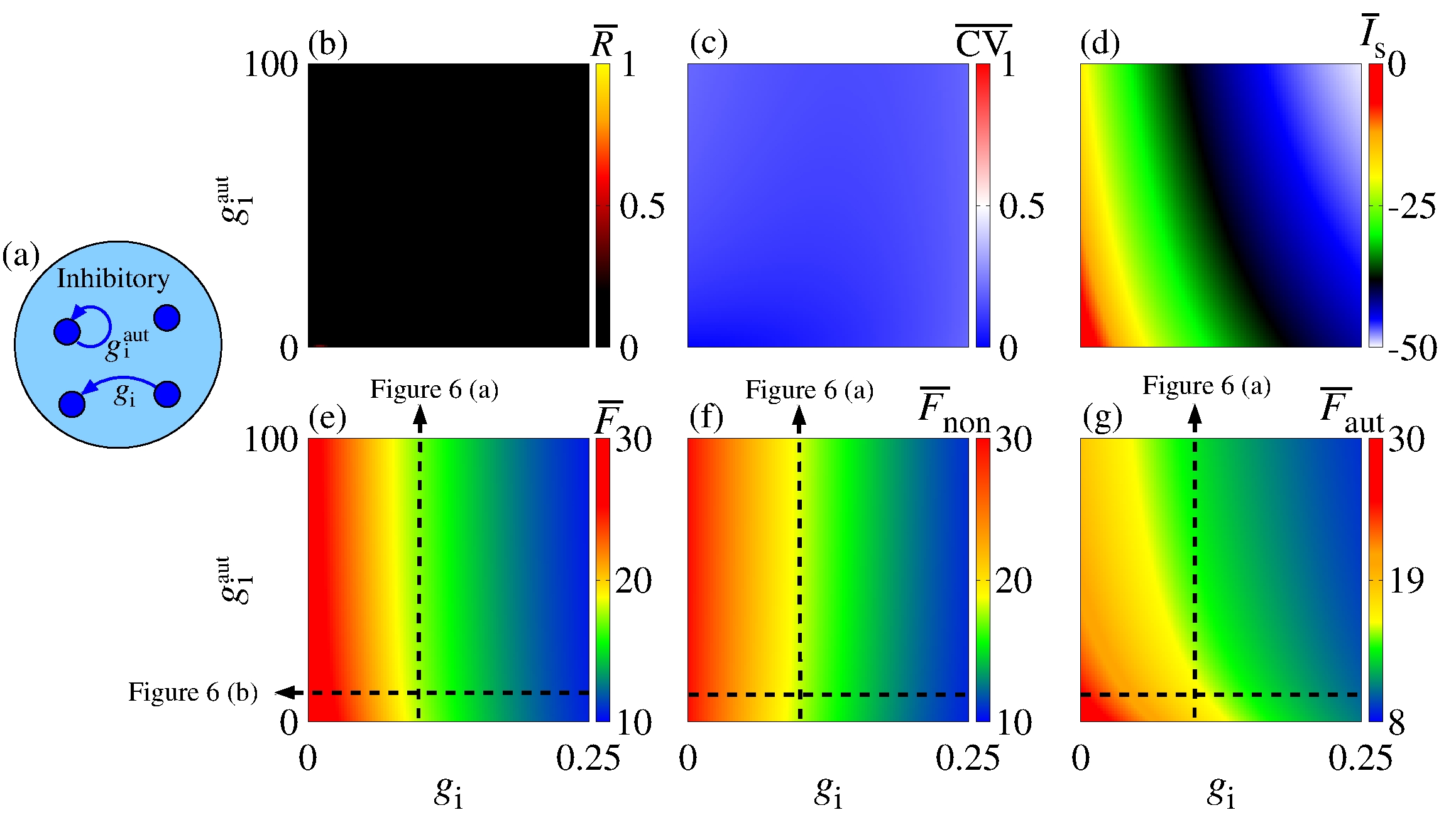}
	\caption{(Colour online) (a) Schematic representation of an inhibitory neural population connected with inhibitory synapses and autapses. Parameter space $g_{\rm i}\times g_{\rm i}^{\rm aut}$, where the colour bars encode the values of \mbox{(b) $\overline{R}$}, (c) $\overline{\rm CV}$, (d) $\overline{I}_{\rm s}$, (e) $\overline{F}$, (f) $\overline{F}_{\rm non}$ and (g) $\overline{F}_{\rm aut}$. The vertical and horizontal black, dash, lines in panels (e) to (g) are used to vary the corresponding parameters in the computations in panels (a) and (b) in Fig. \ref{fig6}. The closed loop in panel (a) corresponds to an autapse of conductance intensity $g_{\rm i}^{\rm aut}$.}\label{fig5}
\end{figure}

Synaptic inhibition regulates the level of neural activity and can prevent hyper excitability \cite{Frohlich2016}. Studies have shown that neural networks can exhibit synchronous activities due to inhibitory synapses \cite{Vreeswijk1994,Elson2002,Franovic2010,Chauhan2018}. Here, we analyse the influence of inhibitory synapses and autapses by varying $g_{\rm i}$ and $g_{\rm i}^{\rm aut}$, as shown in Fig. \ref{fig5}(a). Panel (b) in the same figure shows that inhibitory synapses and autapses do not give rise to the increase of neural synchronisation in the network. Actually, neural synchronisation due to inhibition is possible when it is considered together with other mechanisms related to neural interactions \cite{Bartos2002}, e.g. with gap junctions associated to inhibitory synapses \cite{Guo2012,Bou-Flores2000,Beirlein2000,Kopell2004,Bartos2007,Pfeuty2007,Reimbayev2017}.

In our simulations, we do not observe that inhibitory interactions promote synchronisation in the network. Although this is not surprising, it helps to identify the role of inhibitory autapses in neural synchronisation. Figure \ref{fig5}(c) shows that there is no change from spike to burst patterns, either. In Fig. \ref{fig5}(d), we verify that both inhibitory synapses and autapses increase the intensity of the mean negative synaptic current.

\begin{figure}[hbt]
	\centering
	\includegraphics[scale=0.08]{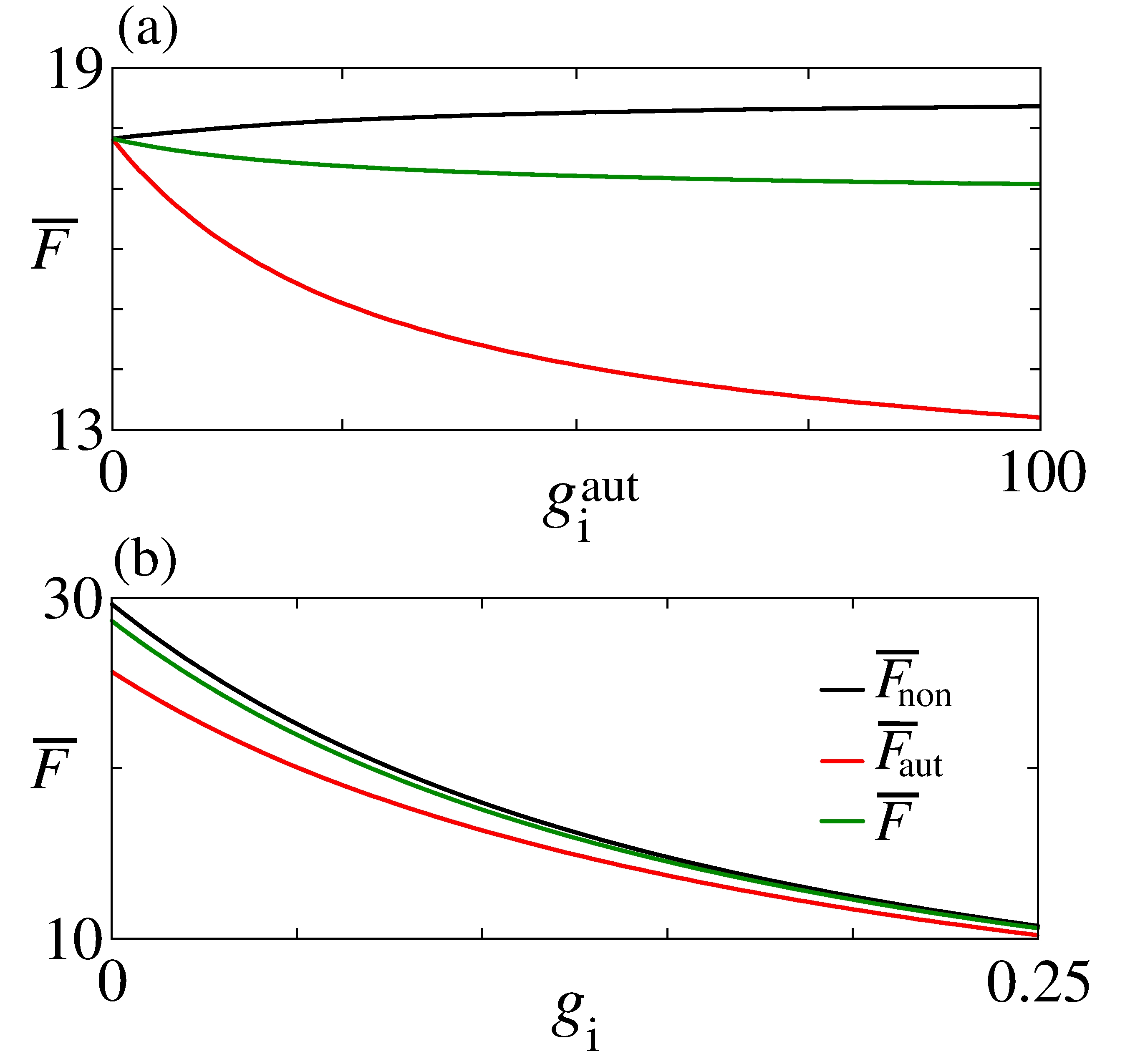}
	\caption{(Colour online) Plot of $\overline{F}_{\rm non}$ (black line), $\overline{F}_{\rm aut}$ (red line) and $\overline{F}$ (green line) for (a) $g_{\rm i}=0.1$ nS varying $g_{\rm i}^{\rm aut}$ and (b) $g_{\rm i}^{\rm aut}=10$ nS varying $g_{\rm i}$, indicated in Fig. \ref{fig5} by the black, dash, lines.}\label{fig6}
\end{figure}

In panel (e) in Fig. \ref{fig5}, we see that inhibitory synapses contribute to the decrease of $\overline{F}$, while panels (f) and (g) in show the mean firing rate for non-autaptic neurons, i.e. neurons without autapses ($\overline{F}_{\rm non}$) and for autaptic neurons ($\overline{F}_{\rm aut}$), respectively. The autapses reduce the firing-rate of the autaptic neurons, what can lead to an increase of the firing rate of the non-autaptic neurons. This can be better observed in Fig. \ref{fig6}(a), which shows the values of $\overline{F}_{\rm non}$, $\overline{F}_{\rm aut}$ and $\overline{F}$ as a function of $g_{\rm i}^{\rm aut}$ for $g_{\rm i}=0.1$ nS. Figure \ref{fig6}(b) shows the mean firing rates as a function of $g_{\rm i}$ for $g_{\rm i}^{\rm aut}=10$ nS. The neurons with inhibitory autapses have lower firing rates.

\subsection{Network with a mix of excitatory and inhibitory neurons}

\begin{figure}[hbt]
	\centering
	\includegraphics[scale=0.18]{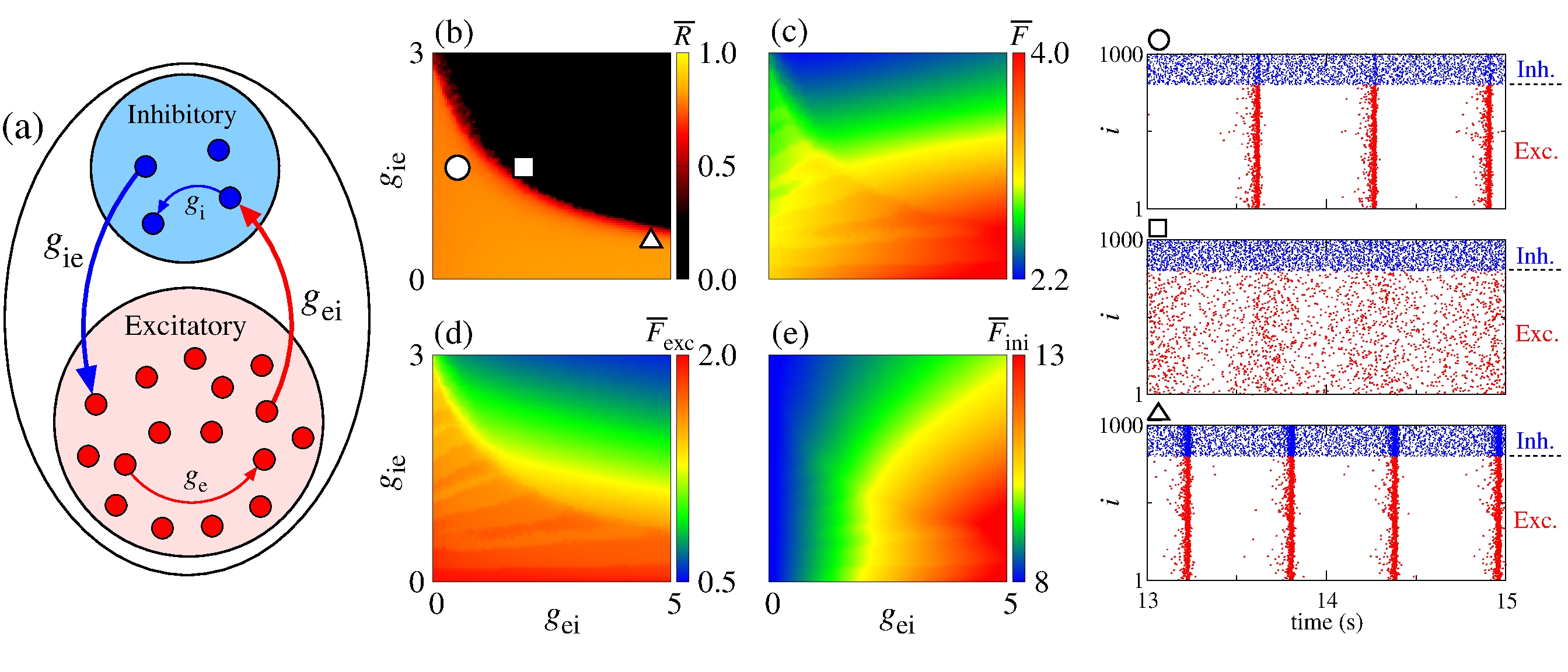}
	\caption{(Colour online) (a) Schematic representation of a neural network with a mix of excitatory and inhibitory neurons without autapses. Parameter spaces $g_{\rm ei}\times g_{\rm ie}$ for $g_{\rm e}=0.5$ nS and $g_{\rm i}=2$ nS, where the colour bars correspond to (b) $\overline{R}$, (c) $\overline{F}$, (d) $\overline{F}_{\rm exc}$ and (e) $\overline{F}_{\rm inh}$. The circle, square and triangle symbols in (b) represent the values of the parameters considered in the computation of the raster plots shown in the right side. The blue and red points in the raster plots indicate the firing of the inhibitory and excitatory neurons over time, respectively.}\label{fig7}
\end{figure}

Desynchronous neural activities in balanced excitatory/inhibitory regimes have been reported in \cite{Borges2020,Ostojic2014}. Based on these results, here we study different combinations of $g_{\rm e}$, $g_{\rm i}$, $g_{\rm e}^{\rm aut}$ and $g_{\rm i}^{\rm aut}$ values in the parameter space $g_{\rm ei}\times g_{\rm ie}$ (see Fig. \ref{fig7}). The existence of synchronous and desynchronous activities depend on the values of these parameters which are related to the conductances. We focus on a set of parameters for which synchronous activities appear. Firstly, we consider $g_{\rm e}=0.5$ nS and $g_{\rm i}=2$ nS in a neural network without autaptic connections.

Figure \ref{fig7}(a) shows a schematic representation of excitatory (red circles) and inhibitory (blue circles) neurons, where $g_{\rm ei}$ ($g_{\rm ie}$) correspond to the conductance from excitatory to inhibitory (from inhibitory to excitatory) neurons in the absence of autapses. Panel (b) presents the mean order parameter ($\overline{R}$) and the circle, square and triangle symbols indicate the values of the parameters considered in the computation of the raster plots shown in the right hand-side. The values of the conductances used to compute the raster plots are given by $g_{\rm ei}=0.5$ nS and $g_{\rm ie}=1.5$ nS for the circle, $g_{\rm ei}=1.8$ nS and $g_{\rm ie}=1.5$ nS for the square, and $g_{\rm ei}=4.5$ nS and $g_{\rm ie}=0.5$ nS for the triangle symbols. The blue and red points in the raster plots represent the firing of the inhibitory and excitatory neurons over time, respectively. Kada et al. \cite{Kada2016} reported that synchronisation can be suppressed by means of inhibitory to excitatory or excitatory to inhibitory connection heterogeneity. Here, we observe that a minimal interaction between the excitatory and inhibitory neurons is required to suppress high synchronous patterns. In Fig. \ref{fig7}(c), we verify that $\overline{F}$ decreases when $g_{\rm ie}$ increases. Panels (d) and (e) show that $\overline{F}_{\rm exc}$ and $\overline{F}_{\rm ini}$ can decrease when $g_{\rm ie}$ increases. In addition, $\overline{F}_{\rm exc}$ decreases and $\overline{F}_{\rm ini}$ increases when $g_{\rm ei}$ increases. When the neural populations are uncoupled ($g_{\rm ei}=g_{\rm ie}=0$), the firing rate difference in the excitatory and inhibitory neurons are mainly due to the adaptation properties of these cells.

\begin{figure}[hbt]
	\centering
	\includegraphics[scale=0.19]{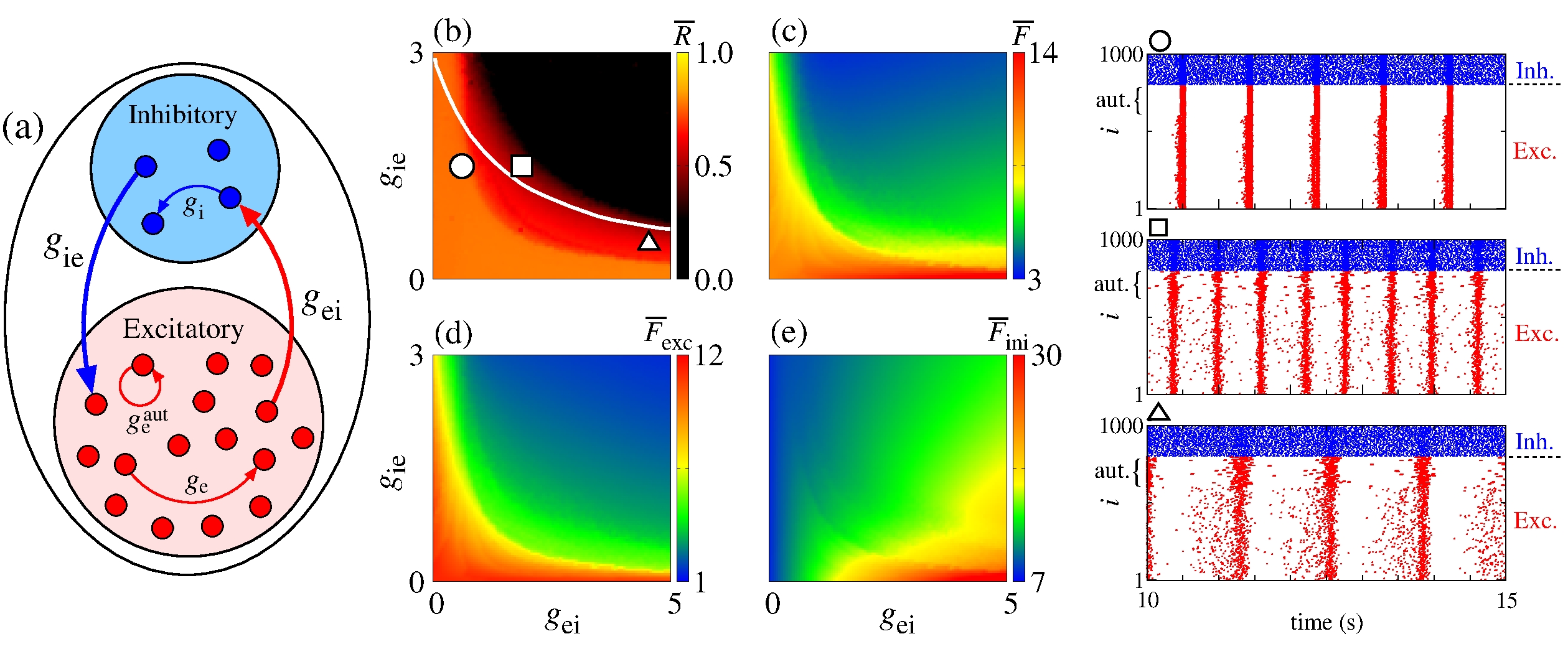}
	\caption{(Colour online) (a) Schematic representation of a neural network with a mix of excitatory and inhibitory neurons with excitatory autapses. Parameter spaces $g_{\rm ei}\times g_{\rm ie}$ for $g_{\rm e}=0.5$ nS, $g_{\rm i}=2$ nS and $g_{\rm e}^{\rm aut}=30$ nS, where the colour bars correspond to (b) $\overline{R}$, (c) $\overline{F}$, (d) $\overline{F}_{\rm exc}$ and (e) $\overline{F}_{\rm inh}$. The circle, square and triangle symbols in (b) represent the values of the parameters considered in the computation of the raster plots shown in the right side. The blue and red points in the raster plots indicate the firing of the inhibitory and excitatory neurons over time, respectively. The curly brackets in the upper left corner of the plots denote the autaptic neurons considered.}\label{fig8}
\end{figure}

Figure \ref{fig8}(a) shows a schematic representation of a network with a mix of excitatory and inhibitory neurons in the presence of excitatory autapses. In panel (b), we present the parameter space $g_{\rm ei}\times g_{\rm ie}$ for $g_{\rm e}^{\rm aut}=30$ nS, where the colour bar corresponds to $\overline{R}$. The white solid line in the parameter space indicates the transition from desynchronous to synchronous behaviour in the network without excitatory autaptic conductance ($g_{\rm e}^{\rm aut}=0$), as shown in Fig. \ref{fig7}(b). The raster plots in the right hand-side of the figure are computed using the values of the parameters indicated by the circle, square and triangle symbols in \mbox{panel (b)}. In panels (c) to (e), we see that excitatory autapses can increase the firing rate of all neurons, changing the mean firing rate dependence on $g_{\rm ei}$ and $g_{\rm ie}$. 

\section{Conclusions}\label{sec_conclusions}

In this paper, we investigated the influence of autapses on neural synchronisation in networks of coupled adaptive exponential integrate-and-fire neurons. Depending on the parameters of the system, the AEIF model exhibits spike or burst activity. In our simulations, we considered neurons randomly connected with chemical synapses in the absence or presence of autapses.

We verified that the type of synaptic connectivity plays a different role in the dynamics in the neural network, especially with regard to synchronisation. It has been reported that excitatory synapses promote synchronisation and firing pattern transitions. In our simulations, we found that excitatory autapses can generate firing pattern transitions for low excitatory synaptic conductances. The excitatory autaptic connections can promote desynchronisation of all neurons or only of the autaptic ones in a network with neurons initially synchronised. The excitatory autapses can also increase the firing rate of all neurons. In a network with only inhibitory synapses, we did not observe inhibitory synapses and autapses promoting synchronisation. We saw a reduction and increase of the firing rate of the autaptic and non-autaptic neurons, respectively, due to inhibitory autapses.

Finally, in a network with a mix of excitatory and inhibitory neurons, we saw that the interactions among the populations are essential to avoid high synchronous behaviour. The excitatory to inhibitory synaptic connectivities promote the increase (decrease) of the firing rate of the inhibitory (excitatory) populations. On the other hand, the inhibitory to excitatory synaptic connectivities give rise to the decrease of the firing rate of both populations. We observed that the excitatory autapses can reduce the synchronous activities, as well as induce neural synchronisation. For small conductances, excitatory autapses can not change synchronisation significantly. Consequently, our results provide evidence on the synchronous and desynchronous activities that emerge in random neural networks with chemical, inhibitory and excitatory, synapses where some neurons are equipped with autapses.

In a more general context, the role of network structure upon synchronicity in networks with delayed coupling and delayed feedback was studied, and very general classifications of the network topology for large delay were given by Flunkert et al. \cite{Flunkert2010,Flunkert2013}, e.g., it was shown that adding time-delayed feedback loops to a unidirectionally coupled ring enables stabilisation of the chaotic synchronisation, since it changes the network class. We believe that the absence or presence of autapses has similar effects upon synchronisation. In future works, we plan to compute the master stability function of networks with autapses to compare with the stability of synchronisation in delay-coupled networks.


\section*{Conflict of Interest Statement}

The authors declare that there is no conflict of interest.


\section*{Author Contributions}
All authors discussed the results and contributed to the final version of the manuscript.


\section*{Funding}

This study was possible by partial financial support provided by the following Brazilian government agencies: Funda\c c\~ao Arauc\'aria, National Council for Scientific and Technological Development (CNPq), Coordena\c c\~ao de Aperfei\c coamento de Pessoal de N\'ivel Superior-Brasil (CAPES) and S\~ao Paulo Research Foundation (FAPESP) (2020/04624-2). We also wish to thank Newton Fund,  IRTG 1740/TRP 2015/50122-0 funded by DFG/FAPESP and the RF Government Grant 075-15-2019-1885. Support from Russian Ministry of Science and Education ``Digital biodesign and personalised healthcare''.


\end{document}